\documentclass[aps,pra,showpacs,showkeys,onecolumn,notitlepage,groupedaddress]{revtex4-2}
\usepackage{comment}
\setcitestyle{numbers,square}
\usepackage{array}
\usepackage{amsmath}
\usepackage{units}
\usepackage{graphicx}
\usepackage{braket}
\usepackage{xfrac}
\usepackage{enumitem}
\usepackage[normalem]{ulem}
\usepackage{booktabs}
\usepackage{tabularx}
\usepackage{tikz}
\usepackage{multirow, bigstrut}
\usepackage{hhline}
\usepackage{color,soul}

\usepackage{times}

\usetikzlibrary{shapes.geometric, arrows, matrix, positioning}
\usetikzlibrary{fit}
\tikzset{%
highlight/.style={rectangle,rounded corners,draw,
fill opacity=0.5,thick,inner sep=0pt}
}

\usepackage{hyperref}
\hypersetup{
colorlinks=true,
linkcolor=cyan,
filecolor=magenta, 
urlcolor=blue,
}

\begin{document} 

\title{Navigating Hype, Interdisciplinary Collaboration, and Industry Partnerships in Quantum Information Science and Technology: Perspectives from Leading Quantum Educators}


\author{Liam Doyle\thanks{Corresponding author: lcd23@pitt.edu}, Fargol Seifollahi and Chandralekha Singh}

\affiliation{
Department of Physics and Astronomy, University of Pittsburgh, Pittsburgh, PA, 15260 USA}

\begin{abstract}

The rapid advancement of quantum information science and technology (QIST) has generated significant attention from people in academia and industry as well as the public. Recent advances in QIST have led to both opportunities and challenges for students and researchers who are curious about the true potential of the field amid hype, considering whether their skills are commensurate with what the field needs, and contemplating how collaborating with industries may impact their research including their students' ability to publish their research. This qualitative study presents perspectives from leading quantum researchers who are educators on three critical aspects shaping QIST's development:
(1) the impact of hype in the field and strategies for managing expectations, (2) approaches to creating conducive environments that attract students and established scientists and engineers from non-physics disciplines, and (3) effective models for fostering university-industry partnerships that can be valuable for students and researchers alike. These aspects, along with several interconnected challenges that the QIST community faces, were explored through in-depth interviews with quantum educators. 
Our findings reveal nuanced perspectives on managing the hype cycle, with experts acknowledging both its benefits in attracting talent and funding, and its risks in creating unrealistic expectations. Regarding greater interdisciplinary engagement and attracting more non-physicists to QIST, educators emphasized the need to recognize and leverage existing expertise from fields such as computer science, materials science, and engineering, while developing tailored educational pathways that meet diverse student backgrounds to prepare them for the QIST workforce. On university-industry partnerships, respondents highlighted successful models, some of them with specific focus on student development while noting persistent challenges around intellectual property, confidentiality, and differing organizational goals. These insights provide valuable guidance for educators, policymakers, and industry leaders working to build a sustainable quantum workforce while maintaining realistic expectations about the field's trajectory as we celebrate the International Year of Quantum Science and Technology.
\end{abstract}

\maketitle



\section{Introduction and Theoretical Framework}

The past few decades have witnessed remarkable growth in the interdisciplinary field of quantum information science and technology (QIST), which promises revolutionary changes in computing, communication, and sensing by harnessing quantum superposition and entanglement \cite{european,raymer2019,flagship,alexeev2021quantum,divincenzo,lloyd,daley2022,altmansimulation,logicalqubit,advantage}. This rapid progress \cite{aaronson2013, levymrs, qkd2017,trapped,semiconductorqubit,2020superconducting,circuitqed,neutralatom,kimble,awschalom,photon,entanglementphysics} has attracted substantial investment from both public and private sectors, generated significant media attention, and created new educational and workforce development challenges \cite{fox2020cu, singhasfaw2021pt,meyer2022cu, asfaw2022ieee,muller2023prperworkforce, qtmerzeletal,bitzenbauer3,weissmanphysics,nvcenter1,nvcenter2,jeremytpt,jeremyajp,Benlarmorajp2025}.

As QIST strives to transition from fundamental research to potential commercial applications, the field faces several critical challenges that extend beyond technical hurdles. For example, the considerable media attention and investment have created hype \cite{hype1} that, while attracting talent and resources, may also lead to unrealistic expectations about timelines and capabilities. Managing these expectations while maintaining enthusiasm for the field's genuine potential requires careful navigation by educators \cite{kohnle2013,rodriguez2020designing,goorney2024framework,bungum2022quantum, michelini2023research, singh2015review,bondani,donhauser2024empirical,chhabra2023undergraduate,marshman2015} and researchers.
Moreover, the inherently interdisciplinary nature of QIST \cite{Preskill,nielsen2010quantum,mermin,QISresource,galvez2014resource,raymerbook,wong,dancing,quantumoptics} demands expertise from multiple fields including physics, computer science, engineering, materials science, and chemistry. However, the field has historically been dominated by physicists, potentially limiting the diversity of approaches and perspectives needed to overcome current technical challenges. Creating pathways for meaningful participation by researchers as well as students \cite{beckphoton,singh2007comp,marshman2016ejpphoton,Kohnle_2017,devore2020qkd,maries2020mzidouble,kiko,singh2022tpt, qtmerzel, hennig2024new, qtbrang, qthellstern, qtgoorney,qtmeyercu,qtsun,hubloch,hucomputing,ghimire2025epj,lopez2020encrypt,  michelini2022, schalkers2024explaining,justicemathphysics} from non-physics backgrounds is important for the field's continued growth.
Furthermore, effective partnerships between universities and industry may particularly benefit students who are interested in non-academic careers in QIST while advancing research and making the transition from academic research to practical applications smooth. These collaborations face challenges including intellectual property concerns, differing organizational cultures and timelines, and balancing openness in research with confidentiality requirements. Understanding how to foster productive partnerships while preserving the benefits of academic flexibility for open scientific exchange is crucial for innovation in QIST.

This paper is part of our investigation on perspectives from leading quantum researchers, who are educators, on different aspects of QIST \cite{ghimire2025reflections,kashyap2025strategies,fargol,liam,liam2} and presents three interconnected challenges. By capturing expert views on managing hype, fostering interdisciplinary collaboration, and building productive university-industry partnerships, we aim to provide educators and other stakeholders 
insights that can guide the development of a sustainable and inclusive quantum ecosystem. These perspectives are particularly valuable as they come from individuals who not only conduct cutting-edge research, but also shape the next generation of quantum scientists and engineers through their educational activities. Our investigation builds on previous work examining misinformation in QIST \cite{kashyap2025strategies}, strategies for diversifying the quantum workforce \cite{ghimire2025reflections}, QIST related courses and curricula \cite{fargol}, suggestions on building blocks to develop a framework for QIST education \cite{liam2}, and the current state and future prospects of quantum technologies \cite{liam}. This study focuses specifically on the sociological and organizational challenges that may ultimately prove as important as technical hurdles in determining participation of students from diverse disciplines, and the broader trajectory of the second quantum revolution.

This investigation draws upon multiple theoretical perspectives from science and technology studies and studies that shed light on the complex sociotechnical dynamics \cite{framework1,framework2} shaping QIST's development. Our analytical approach integrates three complementary frameworks that collectively illuminate the challenges examined in this study.

The Social Construction of Technology (SCOT) \cite{bijker1987social,pinch1987social} framework emphasizes that technologies do not develop according to an inevitable logic but are shaped by the interpretations, expectations, and actions of relevant social groups. For our first research question focused on understanding the role of hype in QIST development, this perspective is crucial. Different stakeholder groups, e.g., scientists, students, investors, business leaders, venture capitalists, policymakers, media and the public construct different meanings of quantum technologies based on their interests, prior knowledge as well as sources of knowledge (including whether those sources are vetted), timelines, and other criteria. We complement SCOT with insights from innovation studies on hype cycles \cite{hypecycle,fenn2008hype,brown2003hope}, which recognize that emerging technologies often experience periods of inflated expectations followed by disillusionment before achieving more realistic assessments of their potential. 

For our second research question on interdisciplinary research in QIST, Gieryn's concept of boundary work \cite{gieryn1983boundary,gieryn1999cultural} provides a lens for understanding how scientific disciplines maintain their distinctiveness while also enabling collaboration across disciplinary boundaries. Although the framework of boundary work has been applied over the last few decades to boundaries between many different disciplines including how these disciplinary boundaries are maintained or crossed, Gieryn described boundary work between science and non-science proposing that boundary work involves the ``attribution of selected characteristics to [an] institution of science 
(i.e., to its practitioners, methods, stock of knowledge, values and work organization) 
 for purposes of constructing a social boundary that distinguishes some intellectual activities as `non-science' [outside that boundary]" \cite{gieryn1983boundary,gieryn1999cultural}. In the context of QIST, instead of non-scientists, this boundary work framework helps explain the challenges non-physicists face when entering QIST, not merely as technical barriers but as socially constructed boundaries that define legitimate participation in the field as non-physicists. 

For our third research question, Etzkowitz and Leydesdorff's Triple Helix model \cite{etzkowitz2000triple,leydesdorff2000triple} conceptualizes innovation as emerging from the dynamic interactions between universities, industry and government. The model identifies universities as generating fundamental knowledge, industry developing applications for market needs, and government providing regulatory frameworks and funding. This framework \cite{etzkowitz2000triple,leydesdorff2000triple} is particularly relevant for understanding university-industry partnerships in QIST, as it recognizes that effective innovation requires not just knowledge transfer but the development of thoughtful organizational norms and shared problem-solving approaches. 

Integrating these frameworks, we conceptualize QIST as a sociotechnical system \cite{framework1,framework2} in which technological development, disciplinary boundaries, institutional relationships and cultural narratives surrounding different types of disciplines and institutions co-evolve. Thus, the challenges identified in our research focusing on the role of hype and how to manage it, fostering interdisciplinary collaboration and building university-industry partnerships are not separate issues but interconnected sociological and organizational aspects of the development of QIST as an ecosystem.

\section{Methodology}

In this investigation, we conducted individual interviews lasting 1 hour-1.5 hours with 13 quantum educators, who are leading researchers in QIST. All educators contacted for interviews were 
well-known quantum researchers with experience in teaching college-level quantum courses, with over a decade of experience in QIST-related research (including Ph.D.). The educators were selected based on their recognized expertise in the field and professional familiarity with one of the authors who knew all but three of them. In these interviews, we asked them many different types of questions. We discussed their views on misinformation \cite{kashyap2025strategies}, how to diversify \cite{ghimire2025reflections} QIST, courses and curricula related to QIST \cite{fargol}, suggestions on building blocks to develop a framework for QIST education, e.g., by finding a common language and appropriate balance of abstraction and physical details \cite{liam2}, and current state and future prospects of quantum technologies \cite{liam}. This paper in the series focuses on quantum educators' views on questions related to hype in QIST and how it is affecting the field, strategies for spawning environments so that more scientists and engineers from non-physics disciplines get involved in QIST, and their views on university-industry partnership for interdisciplinary QIST research and training of students and associated hurdles:

\textbf{Q1. Hype Management in QIST:} What are your thoughts on hype in QIST and how is hype affecting the field?

\textbf{Q2. Expanding QIST Participation:} How can we create environments so that more scientists and engineers from disciplines other than physics are involved in the interdisciplinary QIST research?

\textbf{Q3. University-Industry Partnerships:} What are your thoughts on ways to foster and manage university-industry partnerships for innovation in QIST? 

The interviews were conducted via Zoom in a conversational manner. All interviews were recorded and transcribed automatically. Transcription errors were corrected by listening to the recordings. Repeated words, ``you know'', ``like'', ``sort of'' and other similar common filler words/phrases that participants used in conversations were removed from the transcriptions for clarity.

The researchers used a hybrid inductive-deductive thematic analysis approach to organize and interpret the interview data \cite{braun2006using, braun2021thematic, proudfoot2023inductive,saraswati}. More specifically, the research questions themselves served as pre-defined overarching themes grounded in interview questions. 
The first-round coding and organization of the data into these themes was guided using structural coding, which is a holistic approach based on the answers to the questions asked from the quantum educators \cite{mclellan2008,saldana2021coding,hedlund2013overview}.

Following this initial round of coding to organize responses by themes, the researchers individually investigated patterns in data within each research question to identify subthemes, or what we here refer to as common codes.  After multiple discussions and iterations of finer-level coding (based on smaller chunks of text or sentences), the researchers converged on the final subthemes. These were ultimately presented as common codes nested within each research question (theme).

\begin{figure}
    \centering
    \includegraphics[width=0.5\linewidth]{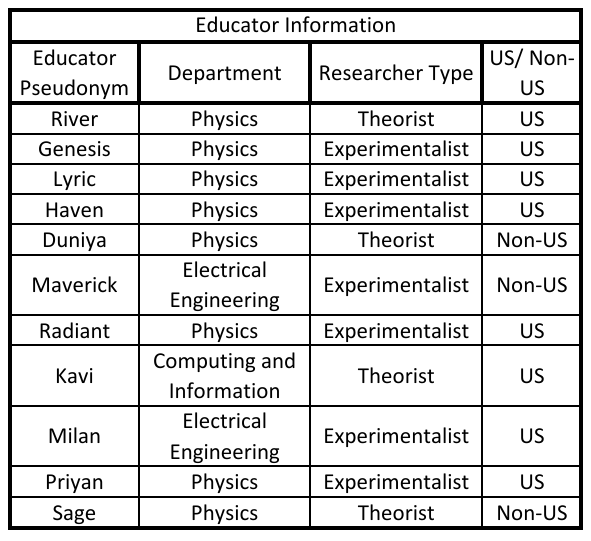}
    \caption{This figure provides information on the educators, their departments, if they are theorists or experimentalists, and if they are based in the US or not, to give context to their statements (although responses of the US and non-US educators were similar).}
    \label{fig:educator}
\end{figure}

Although 13 quantum educators were interviewed, here we only focus on responses of 11 of them, who were all educators at large research universities in the US. These include educators whose responses provided the most valuable insights and their responses were directly relevant to the research questions asked, since not all educators answered every question and the length of their responses varied greatly.  All the educators whose views are present in the current paper were affiliated with research-intensive institutions that offer PhD programs in relevant fields. Additional information about the 11 educators is provided in Figure \ref{fig:educator}. The educators who are not from the US are from countries that are considered western with majority White populations. All educator names used are pseudonyms.

We note that while several educators elaborated on specific QIST courses and curricula in response to questions about interdisciplinary participation (portions of which are mentioned here), a detailed analysis of educational content is not within the scope of this paper and has been presented in a previous study \cite{fargol} focused on QIST courses and curricula.

\section{Results}

We now discuss the results for each of the research questions organized by the common codes that emerged from quantum educator responses. Although there are overlaps between different codes, these were the codes that all researchers, via multiple discussions and iterations, agreed were the most appropriate based upon the patterns of educator responses.

\subsection{Q1: What are your thoughts on hype in QIST and how is hype affecting the field?}

\begin{figure}
    \centering
    \includegraphics[width=\linewidth]{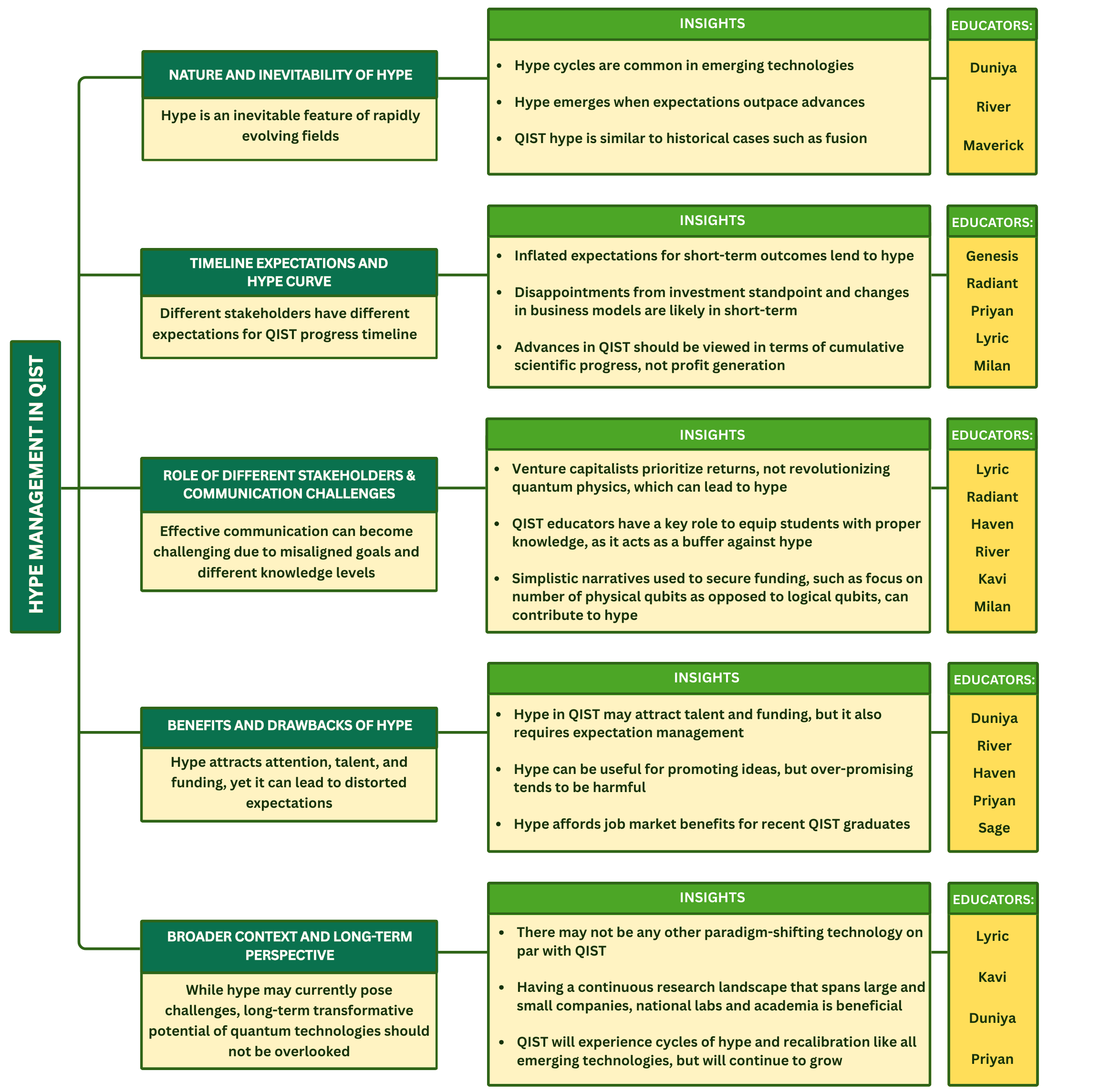}
    \caption{Common codes and insights of educators in response to hype in QIST.}
    \label{fig:RQ1}
\end{figure}

Hype was generally regarded as an unavoidable feature of a field such as QIST that rapidly evolves with potential for unimaginable technological breakthroughs. The educators emphasized how the dynamics of the hype can be closely related to different expectations from different stakeholders, both in terms of timelines for progress and perceptions of current state of the field. Their views also pointed to communication challenges between different stakeholders, due to various reasons such as different goals and knowledge base. While the educators acknowledged that there are both benefits and drawbacks to hype in this field, they were broadly optimistic about the long-term potential of QIST. The educators' responses related to this research question were categorized under the following codes: ``Nature and inevitability of hype", ``Timeline expectations and the hype curve", ``Role of different stakeholders and communication challenges", ``Benefits and drawbacks of hype" and ``Broader context and long-term perspective". The insights provided under each of these categories, as well as educators who mentioned them, are summarized in Figure \ref{fig:RQ1}. Below, we present detailed reflections of educators on these topics.

\subsubsection{Nature and Inevitability of Hype}

Some quantum educators emphasized that hype is guaranteed in a rapidly-growing field that has transformative potential for future technologies.

For example, Professor Duniya stressed that hype cycles are common in emerging technologies saying, ``Essentially, every major emerging technology goes through this cycle of hype and it's very difficult to avoid it. A lot of it is driven by the way the business works, and as soon as you get things like venture capital investment [involved, hype is bound to happen]. It's really important for companies to get the bigger picture out there [but] often that gets oversimplified, and it generates a lot of interest very quickly, in a way that doesn't manage expectations properly. In a way, that doesn't often convey the key messages of what the real potential of quantum technology is properly.''

Professor River thought that hype is inevitable in fields in which societal expectations about outcomes outpace actual technological advances. They said, ``hype has to do partly with what society thinks, in some sense how much people are pushing at it, and how quickly its promise gets realized. Just to give an analogy...fusion [for producing energy in reactors] \cite{fusion1} has got its ways up and down...it's just that it was hyped up for decades before we got our preliminary results...".

Professor Maverick recollected interesting experiences related to hype in QIST. They reflected on a conference they attended saying, ``you could call it a hype fest. If you come from the pure academic point of view, you will get physical pain in your body at the sort of stuff that some people say on stage. But someone at that conference said something that really stuck with me. They said, you see, hype is necessary to create the fear of missing out among investors. Hype is a business necessity." They continued,  ``I'll tell you another story...because it really stuck with me. A few years ago, I was in a discussion panel on quantum computing and quantum technology, and I was invited there as an academic expert. There were some people from government, some people from industry; and there was this guy: He was the CEO of a major bank...he was going on and on about, `oh, yeah, quantum computers, they're already here. They're going to revolutionize the way we do finance'. I was kind of biting my tongue at first, and then saying, `look, it's actually not that simple, the progress in it still needs to be made'...It was obvious we did not get along with each other...[then] this guy turns around to me and says, `you don't understand how the money gets made, do you?' And that one really stuck with me...I know how to write the Schr\"{o}dinger equation and solve it. I know how to control spins in silicon...[but] who am I to tell these people what to do with their money?''

\subsubsection{Timeline Expectations and the Hype Curve}

Some quantum educators focused on how differences in expectations regarding the timeline for the development of quantum technologies can lead to hype. They also provided interesting views on where quantum technologies are on the hype curve.

For example, Professor Genesis reflected on hype by stressing that it is fueled by people's incorrect perceptions of how quickly practical applications will be achieved. They said, ``I think we underestimate long-term progress, and we overestimate short-term rate of change. We think things are going to change very quickly. They don't. But over the longer term, there will be a big change, is always my feeling...''. Professor Genesis compared the situation in QIST with artificial intelligence (AI) \cite{aireview,ai2} and noted, ``We are probably due for a quantum winter or the valley at some point because just from the fact that some of these applications are still a few years away, there's inevitably going to be investor disappointment that they don't have the next big thing already done." They believed that the ``winter" from the investment standpoint will happen in the near future, and compared this to AI hype: ``then it'll be like AI [artificial intelligence] 20 years ago. There was a lot of hype about AI...and then that went into a winter for a long time, but people kept working on it and eventually, they've now come out with all these new things...I foresee that quantum computing will probably follow something similar.''

Professor Radiant started their reflection by summarizing their views on the current landscape: ``so now we have a very large number of companies doing quantum software and quantum computing and quantum communication...It's a great growth industry and the stocks are doing very well for most of them. Some of them have crashed and burned.'' Forecasting what would happen in the future, Professor Radiant added, ``I think we're at the point now in the field probably [where] it is going to contract a little bit in the next 5 years. As companies find out they're not making money from this, and the investors pull out, some companies will go out of business or change their business model. Some of the companies might move away from it after 5 or 10 years because it's not panning out."
They continued, ``But then I think over the next 20 years, scientists will keep working on it. And then, in the long run, most people are very optimistic that really great practical results will come out eventually. But think about the classical computer; it took really 40 years to go from the first microchip to the internet where it really had an impact on people's daily life. And during that time, it was a small impact. It was mostly in high-tech companies, government labs and places like that. So, I think quantum information will be like that. It'll be another 20 years before it really starts to impact people's lives like self-driving cars...the navigating without GPS \cite{gpsnavigation}, and so on.''

Regarding where the quantum computing companies are on the Gartner hype curve \cite{hypecycle}, Professor Priyan said, ``From what I can tell...my gut is that they are not on the downward slope yet".

Professor Milan had different views. They believed that the hype in quantum computing is not as exaggerated as it used to be, saying ``I'm under the impression that it's getting a bit better...[there] are a lot of really serious hardware companies working on quantum computing. I think it's still really hard to motivate why someone would invest into this and it's probably easier for big players like IBM or Google to just find internal funds. Also, I think the whole ecosystem is getting more educated in what it is, and what it can do and what it cannot do, and what's worth the effort or what the potential of each different platform or each different approach to algorithms is. So, I think that way, because it's been around for a bit more, just through the education of the community that things are just a bit less hyped".

Professor Lyric reflected on the hype cycle \cite{hypecycle} saying, ``My take on the hype curve is that instead of showing the hype curve, the curve that looks like there's a big rise and then there's a dip, and then there's like a plateau, you should integrate the hype curve...So a plateau means...you're not static...that means it's actually contributing to a lot of stuff, that the plateau is actually leading to...technologies. So if you integrate even at the dip of the hype curve, you're actually much better off than when you started." They explained how the ``valley of death" is important from the view point of making successful, profit-generating companies. 
They continued, ``But I think that it [quantum computing] is not just about profits. It's about something much bigger! Would you say that quantum mechanics was for profits in the twentieth century?...No...it was about developing and creating a foundation for the field of chemistry, for creating and understanding optics and solid state...''. Professor Lyric preferred not to think about where we are on the hype curve in QIST saying, ``I think this is the wrong way of thinking about what's going on here [in QIST]...It's just kind of like buy, sell, hold kind of thing [in stock market]...Just take a step [back]...and then see how ridiculous it would sound a century ago [when the first quantum revolution was taking place].''

\subsubsection{Role of different Stakeholders and Communication Challenges}

Some quantum educators emphasized that different stakeholders in QIST have different understanding of the current state of the field, and how effective communication is challenging between researchers who have a deep knowledge of the status of the field and outsiders who may be struggling to get a clear picture of the current status, which can contribute to hype. 

For example, Professor Lyric emphasized that different stakeholders may have different levels of understanding, and while communicating with a QIST expert, their levels of focus on details or the amount of attention they pay depends on their different goals, which can further contribute to hype. They said, ``The problem is that you can't talk to a reporter...at any level of technicality, because it's like they don't want to hear about that. They just want to write their article about the quantum computer that's coming or that's already here.'' 

Professor Lyric noted that QIST hype also has to do with the fact that the number of physical qubits as opposed to logical qubits is often emphasized by companies. They said, ``It's easier to say oh, we made 300 qubits. And now we're going to go to 700 qubits. And now we've gone to 1,500 qubits. Yeah, it could be that that's actually needed to make one logical qubit. So, the point is that all of this development is necessary. It's not one versus the other." But in terms of how to secure funding, they continued, ``As far as how to sell to upper management or to the public, or to the people who are giving you money, how do you sell that?... [if you say] we want to make one [logical] qubit...[The management will say], you can't even make one [logical] qubit? It's not a great way to sell what you do if you say I haven't even made one logical qubit so that it can withstand the decoherence over [any reasonable] amount of time. It doesn't sound great. But it's the reality of where we are right now." They noted that ``we're just going to get better and better at making these big windup toys and the problem is that the way that it scales is not that good". Regarding related issue of getting the errors down, they
said, ``[it is] going to be difficult to get the quantum volume that you need to do useful computation, although if we get these errors to go down by even just like a factor of 2, it has a huge impact on the quantum volume...The number of forces that are fighting against you when you go from something...to the next level, it can be [too many and so it is challenging to make rapid progress in making logical qubits]".

Professor Radiant echoed these views, emphasizing that we need to recognize, e.g., that the goals of venture capitalists and scientists vary, and that different timeline of venture capitalists compared to researchers about QIST is related to hype. They stated, ``The goal of most venture capitalists is actually to earn money. It's not to revolutionize quantum physics".

Professor Haven felt that as a QIST researcher who reads original research articles and not press releases or media articles, they can not say much about hype in this discipline. They said, ``it's hard for me to comment on hype, because I [as a quantum computing researcher] don't engage with these announcements like a normal person would, like a person from even another field of science. When someone announces a new vaccine or a new medicine, or a new airliner, there I'm not an expert. There I have to look at hype and judge what it means for me. In this particular area [QIST] where I'm an expert, I don't feel like I'm a very good judge of the hype...I just read what the actual paper was.'' 

Professor Haven further commented on their views on why companies engage in hype, ``They're supporting their business...business is not science. So, there's the claim a scientist and a university make. There's the claim a business makes backed up by a scientific paper. And then there's business talking about business. None of these things is really the same...Yeah, the claims should be justified by science, and they should be reasonable...But it definitely is a different world now, where there are press releases that accompany these scientific results..." They explained how far the disconnect between scientific community and media can go: ``I can't believe whenever a picture of a quantum computer shows up in Forbes or the New York Times. I'm always shocked because in reality these are very small research fields that are very deep in this hole [e.g., quantum computing research]. I'm always personally amused when they have a picture of the wrong kind of quantum computer, like when it's an article about an ion trap \cite{trapped} and they have a picture of an IBM fridge \cite{ibmising,ibmquantum}, or when...[it is] an IBM fridge they're talking about, they have a picture of an ion trap. That shows you the immense level of disconnect." They continued, ``But those layers are not always even determined by scientists. So, what you can do is make sure your scientific paper has claims that are supported by the evidence and are reasonable. Then if you go off and make vast claims to the media, that's not good scientific research. But if you put out a reasonable press release and it gets massively distorted by some media that's trying to report on what's a brand-new scientific result, that's weird.''

Professor River reflected specifically on the issue of hype related to quantum computers breaking RSA encryption using Peter Shor's algorithm \cite{shor1994}. They felt that while QIST researchers use the example of possible consequences of the implementation of Shor's algorithm on a scalable fault-tolerant quantum computer, those who are not involved in QIST research are likely to get confused and not recognize that such quantum computers are not going to be built for at least several decades in the future. They said, ``that's easy to convey to a lay person...that all your bank records could be compromised...so in some sense, I can see it's a nice go-to where you can explain to a lay person certain impacts of quantum computing with very minimal technicality...[for example,] when I try to explain to my cousin, this is my go-to example, because he works with computers, and this is something I can explain to him.''

Professor Kavi emphasized the important role of educators in preparing students well so that they are not misled by hype, ``we should prepare them to have a generic worldview, which is well informed. So, by that what I mean is, they should be able to bust all the hype that is out there. They should know what is hype and what is not hype and that will be first [point to emphasize]. Secondly, they should know the capabilities of quantum information processing." They continued, ``The first [of these points I am making] has to do with the second, which is-if through our courses, we are able to give them preparation and give them a sense of what is actually possible at least as of now and what is not [possible], to have a realistic picture of what this whole technology, this revolution is all about, [that would help]...So yeah, busting the myth is the big thing, I think if somebody has a reasonably good background, at the undergrad level, where they can tell what is possible and what is not, [they would not be misled by hype]...''.

Professor Milan echoed sentiments similar to Professor Kavi's and said that they make it a point to clarify their students' doubts about hype and other related issues, particularly because students may get confused by information from different unvetted sources.

\subsubsection{Benefits and Drawbacks of Hype}

Educators pointed out that there are both advantages and disadvantages of hype in a discipline such as QIST. While hype can attract attention, talent, and funding, it can also lead to distorted expectations.

For example, Professor Duniya emphasized that hype is a double-edged sword saying, ``I think that the hype is both helping us and hurting us". They noted some of its advantages, ``It has the massive positive effect that it keeps a lot of people interested in asking questions about quantum mechanics, quantum information science and quantum technologies. It is great to have people interested in physics, especially younger generations. Next generation of potential physicists are coming through, and anything that gets more people interested in physics is a good thing, especially something that can slightly more democratize physics...And of course, industry is interested in what these technologies might be able to do, and I think that is definitely a good thing." They continued to elaborate on the negative aspect, ``The hype, of course, is often very frustrating especially when you're very excited about what quantum computing might be able to do. But you realize that there's a real need to manage expectations because the things that it's likely to deliver are going to be much further off...''. 

Professor River pointed to the positives of hype in QIST while emphasizing that it is important to clarify the current status of the field so as to manage hype, ``I do think maybe there's some good hype about it in the sense that we definitely want more people getting into this. A few of the technologies are definitely growing that people should be putting money in, but in the form of research money because it's important to fundamental physics and future technology. But it's not one of those things where, if I buy a stock, I expect the stock to go up...[but] where we should put investment into...probably knowing that the payout won't be any time in the near future.''

Professor Haven acknowledged that a certain level of hype is useful for QIST advancement. However, it can be problematic if it leads to unrealistic expectations about what is to come and when some expected product, e.g., a quantum computer that will have a quantum advantage on Shor's algorithm, will become available in the near future. They reflected, ``You need people to pay enough attention to get your money [for your research]...past that it [hype] starts to become harmful because then people start to over promise. [People] start to expect unrealistic things, or people who don't really know how it works are the ones deciding what you have to do next." They believed that there may be more hype in QIST than is useful, and they compared the current state of hype in quantum computing to hype about flying cars and how ``that's not driven by science. That's driven by the need to have a competitive stock offering and have people think that you have a vision moving forward". They continued, ``I think quantum is very much in that place right now where when you ask what some company is doing about quantum, it's not because the science is settled and they know exactly what's going on, it's this more general question of how they address future-proofing their companies.''

Professor Priyan noted that at least one advantage of hype is that ``the federal government has put a lot of money into these things. So that [hype] is what it takes [to provide funding for research]."

Professor Sage reflected on the positive aspects of hype, saying they no longer even have to write letters of recommendation for their graduating students because they are being scooped up by QIST-related companies, ``in some ways it [hype] is great. Everybody gets a job. But in other ways, it's terrible...[but] I'm a great believer that in the long run, people will be able to do amazing things [with quantum technologies]". 

\subsubsection{Broader Context and Long-term Perspective}

Some quantum educators emphasized that while hype is currently posing challenges, it is important to keep the long-term potential of quantum technologies in mind, as their transformative potential will definitely unfold in the long term.

For example, while Professor Lyric agreed that hype is prevalent in QIST, they strongly defended large and small companies being involved in making quantum computers at this time. They said, ``I think you need to coarse-grain and think about the big picture, kind of macroeconomics rather than microeconomics. I would say that there's nothing wrong with having a research landscape that involves large companies, that involves startups, that involves research at universities, national labs." They also noted, ``private investors probably already know that the chance that they're going to make a quick buck off one of the technologies is rather slim, or at least they should know that if they've been talking to the right people. But in some sense, it's an investment. It could be part of a patent portfolio that ends up being [profitable in the future] and [may become part of] some other technology that gets acquired by another company.''

Professor Kavi noted that, while from economic and workforce points of view, hype may be relevant in QIST, it is not relevant from a scientific point of view. They said, ``I think there is no choice of falling down because, to my knowledge, if we give up on quantum, I don't see any on par fundamentally path-breaking, paradigm-shifting technology. Maybe I'm wrong, but I do feel like humankind will continue to strive to use the most up to date physical theory to enhance all our capabilities." Professor Kavi made a distinction between the future of QIST from an economic vs. scientific point of view saying, ``In my view, if we fail in the current attempt at quantum information, maybe some entirely new approach might come about. But from an economic point of view and jobs and stuff, maybe there might be a crash. But from a fundamental science point of view in research, I think we would always continue to pursue [QIST]. But I think when we are talking about the workforce, you care more about the economic side. And so, in that regard, there is probably a crash coming in 10 years' time...''

Reflecting on the long-term perspective of QIST, Professor Duniya pointed to how all emerging technologies go through cycles of hype and recalibration before some of them ultimately succeed. They explained, ``There are going to be cycles of hype and not all the startup companies will survive simply because that's how it works in business in every area, and there will be fluctuations, and things will go up and down. There are too many interesting technology areas. There are too many potential applications for these things...There will be, at some point, a series of startups that don't succeed, and then there will be growth again in different small to medium enterprises and large growing companies. There will be [existing] programs close and there will be new programs open. Isn't that how most emerging technology fields tend to work? So, I don't think it's ever as simple as there being a bubble. There will be recalibrations but I'm optimistic about where things are going to go in the future.''

Professor Priyan acknowledged that their knowledge about hype may be lacking because they do not engage with it much, ``I don't follow hype, politics or financials of these things. I only get what I get at random conferences where I step into random talks. So, I get perhaps a random sampling of things, definitely not a complete picture, so I don't know too much about that". However, they expressed that they did not believe there was any hype in quantum sensing. This is an area in which they are involved, and they believed there should be more attention given to it at this time. They said, ``I think in the quantum sensors, there is no hype even though there is commercialization going on, as post quantum computers. I think there are things that are commercializing, especially around sensors in solid state systems, things like NV centers and related objects in solid state. Those, I think, are becoming commercial objects. At least there are companies now that sell them commercially to other scientific enterprises, and I can see the technology is very close to where [they can] make some magnetometers and other types of sensors, which are based on quantum principles...I don't see a ton of hype around it…And to me it seems like a good idea [that should get more attention]." They continued, ``It's a compact technology. It works with room temperature that can be well-engineered, and it can make a better magnetic field sensor than a lot of things that are out there. So, it has got a lot going for it in terms of that type of quantum sensors; in terms of commercialization [yet, it has not gotten as much attention]".

\subsection{Q2: How can we create environments so that more scientists and engineers from disciplines other than physics are involved in the interdisciplinary QIST research?}

\begin{figure}
    \centering
    \includegraphics[width=\linewidth]{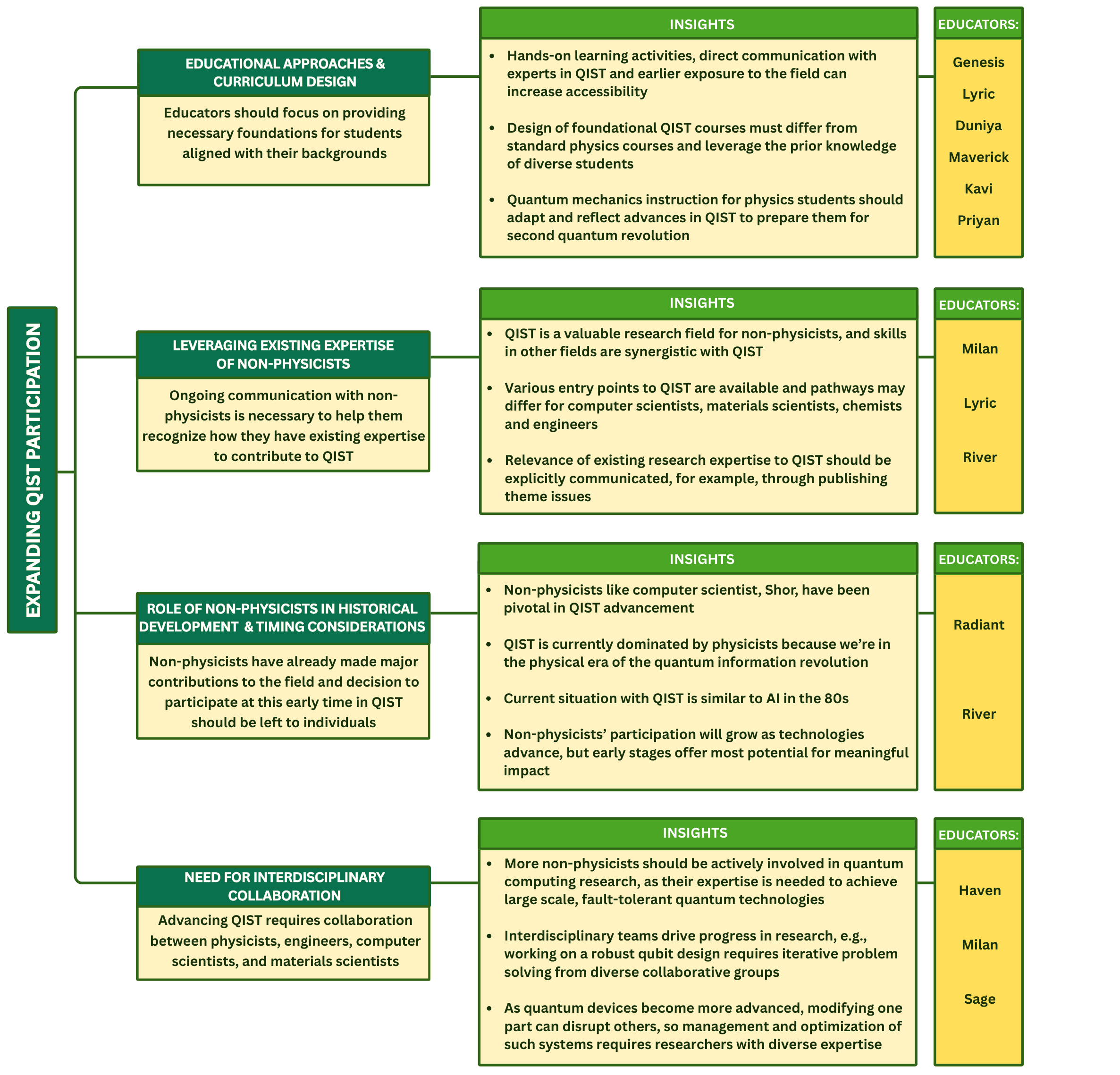}
    \caption{Common codes and insights of educators in response to expanding QIST participation.}
    \label{fig:RQ2}
\end{figure}

In response to this question, some quantum educators focused on the development and instructional design of QIST courses \cite{fargol} that accommodate students from interdisciplinary backgrounds. Several of the educators emphasized leveraging the existing expertise of non-physicists, and the current need for interdisciplinary collaborations. Additionally, some educators emphasized that non-physicists have already played a key role in the historical development of QIST, but since we are very early in the development of quantum technologies, in this ``physical" era, there is a dominance of physicists.
Quantum educators' thoughts related to interdisciplinarity and participation of more researchers from disciplines other than physics in QIST were categorized under the following codes: ``Educational approaches and curriculum design", ``Leveraging existing expertise of non-physicists", ``Role of non-physicists in historical development and timing considerations" and ``Need for interdisciplinary collaboration". The insights provided under each of these categories, as well as educators who mentioned them are summarized in Figure \ref{fig:RQ2}.

\subsubsection{Educational Approaches and Curriculum Design}

Some quantum educators emphasized that the design of QIST courses should take into account the fact that there are students from different disciplinary backgrounds in them and especially focus on providing necessary foundation in this field to non-physicists.

For example, Professor Genesis highlighted an interdisciplinary undergraduate course on foundations of QIST they had recently taught, in which a large fraction of their students were non-physicists. They believed their course was successful in getting non-physicists to learn QIST in an accessible manner; e.g., one strategy they used was giving students Qiskit \cite{qiskit} group coding projects. Professor Genesis also noted that in the future, they plan on having researchers who work on different qubit architectures visit their course, as it could help students from different disciplinary backgrounds get a better feel for the current state of the art in QIST, and where each qubit architecture is in terms of the developmental roadmap at this early stage in the second quantum revolution. They considered such visits to be particularly helpful for students from different disciplines who are looking for more concrete experiences.

Professor Lyric emphasized the value of teaching interdisciplinary QIST foundational courses, while keeping in mind the diversity of student needs and not teaching them as physics courses. They said, ``it's important to come up with some models of the students that you're trying to [attract in this interdisciplinary field]...paint a portrait of 5 or 6 types of students. This person is a physics major and a computer science double major...or this person is a chemistry major and is also interested in this [QIST discipline] based on some article that they read''. Professor Lyric felt that it was important that the courses focused on QIST foundation include ``some special modules that the computer science and physics double major is going to be happy to chew on, and...some projects...that people [with other interests] can use to dive more deeply into this [field] and explore the various pathways that mimic the multi-faceted nature of the field." They further explained, ``If I were structuring a course, I would want to have a very clear understanding of the portraits of these students, who they are, and then make sure that we can match as closely as possible to at least having one or more of these pathways be available to students so that they can benefit, but then also share the [big picture] knowledge [with students from different disciplinary backgrounds who have diverse interests] within our common theme of the course.''

Professor Lyric also stressed on earlier exposure to QIST concepts, and that there are many concepts that educators need to 
``break down in a way that people get exposed to them at an earlier and more formative age so that they know what it means to do these things, and they can aspire to do them from earlier ages, from high school, for example, or even [from] grade school-like I want to be a firefighter, and I want to be this and that, I want to be a quantum computing person...I think that there is a lot of work to be done.''

Professor Duniya was planning to start an interdisciplinary QIST-related degree program at their institution in the near future. They had spent significant effort in communicating about interdisciplinarity issues with their faculty colleagues in engineering and computer science departments, who were also involved in the effort, in addition to colleagues from other universities and disciplines. They emphasized the importance of teaching both undergraduate and graduate level QIST courses to students from interdisciplinary backgrounds in a manner that they did not feel like physics courses to them. They stressed the importance of keeping the mathematics needed in the courses to a lower level than what is usually common in physics courses for students in a particular year in college, and only focusing on finite dimensional Hilbert space. They believed that QIST courses that are aligned with the backgrounds of non-physics students would help them learn foundational QIST concepts well, saying, ``I think, for someone with a much broader mathematical background---someone with a background in computer science and engineering---it often is an easier introduction to these things".

Professor Maverick reflected on the quantum engineering bachelor's degree program which they helped spearhead and emphasized how it was born out of an engineering school to ensure more non-physicists were ready for the quantum workforce~\cite{fargol}. They explained that the idea of this program started from a course they had already taught for a long time, ``It was basically a course of quantum mechanics for engineers using the modern engineered quantum devices as the model system to teach quantum physics. So instead of doing the harmonic oscillator, the hydrogen atom, and all of that, we did quantum dots, Cooper pair boxes for superconducting qubits, SQUIDs, stuff like that". In collaboration with their other faculty colleagues, they expanded this to create some quantum-specific courses and emphasized that ``our graduates are people who are full-blown engineers. If they never want to see a Hamiltonian or Schr\"{o}dinger equation ever again, that's fine". 

Professor Kavi had recently taught a graduate level QIST course, which had a combination of non-physicists and physicists. In this course, they not only invited some early-career researchers with diverse disciplinary backgrounds to their classes to inspire students, they also had students read QIST papers from across disciplines and complete some projects. They were also preparing to teach an undergraduate quantum information course soon which focused on getting students, particularly those from non-physics backgrounds, excited about the potential of quantum technologies and helping them recognize that non-physicists can also play a key role in advancing the field.

Professor Priyan 
acknowledged the need to prepare more non-physicists for the QIST workforce. However, they believed that the preparation of physics students taking quantum mechanics should also adapt and reflect advances in QIST to better prepare them for the second quantum revolution. They said, ``we should have them understand the physics that we have always taught in sort of this more modern language, so at least to some extent, [they will] be able to translate things into the modern language of quantum information, as opposed to just the language of quantum mechanics. Because we definitely teach quantum mechanics to every physics student and we don't necessarily teach [about QIST concepts]". They clarified that "not every quantum mechanics problem should be modeled as gates", yet including some discussion of gates when discussing interacting quantum systems will be useful. 
They considered this type of training to be valuable in part to prepare physics students for the second quantum revolution. 

Professor Priyan added, "Giving students a broad background in applying quantum mechanics to real systems [is important]"
They reflected on the content of available quantum textbooks saying, ``at least when I took…quantum mechanics a while ago…there were not a lot of applications in the quantum mechanics textbook that I learned from. There are maybe a few applications to some basic high energy, say, particle physics. But there's not a lot of talk about how these things relate to not just quantum computers, but all types of systems [related to QIST], quantum communication system, sensing systems, quantum chemistry type things. I think that is lacking a little bit in quantum education. Maybe that needs to be a separate [course]..." They continued, ``But can you do those [discuss QIST concepts in the existing quantum courses]? Can you learn about those things having the basics of having gone through the Schr\"{o}dinger equation, having gone through spin systems, having gone through those independently? Whereas to me, the applications that are traditionally there [in a quantum course], they're just atomic physics type things. You learn about 2 level atoms and radiation, which is good. That's the fundamental thing. That's what I deal with all day long. But how that relates to the rest of the world [like quantum technologies] seems like a useful thing to teach".

Professor Priyan reflected further on the course they were developing and how it will better prepare students for the interdisciplinary QIST field through its focus on applications and engineering aspects. They said, ``I'm on the schedule for doing a graduate quantum optics class in the next semester. So, I'm trying to figure out what to put in there. I'm trying to make it more application-based rather than based on going through the top-down approaches. This is the quantum master equation and...this is how one solves them; which is useful and important information. But mapping systems, mapping the actual behavior of real systems into this math is the more useful and interesting part. So, we'll try and talk more about things like quantum communication protocols and how loss and decoherence affect quantum systems in terms of what you can do with them in terms of making sensors, or just memories." They continued, ``We probably won't talk too much about quantum computers per say, [but rather] how qubit lives [and decoheres] in the real world. That's the topic we'll get into a lot". They gave a concrete example of helping students understand what is signal and noise in these contexts, which are practical issues that are critical for controlling and manipulating systems, important for both physicists and non-physicists. They said, ``If you're the person who makes the sensor and it's up to someone else to sense something, it's their problem what couples into it, not your problem. So, if they're adding noise, then that's signal to you". 
Educator Priyan added that there is a lot one can teach students ``in terms of quantum measurement and everything else. What is your system? What is your measuring device? And precisely defining all those things, and what's the rest of the world, and making those definitions [clear is important because they are] very unclear in real life. Even in doing the textbook examples, it's not always clear what needs to be included or shouldn't [be included in your system]". Thus, Professor Priyan wanted to provide an application-oriented perspective in their graduate quantum optics class that can be valuable for all students regardless of their disciplinary backgrounds.

\subsubsection{Leveraging Existing Expertise of Non-Physicists}

Some educators stressed the importance of leveraging the existing expertise of researchers from non-physics disciplines, who may not realize that their current expertise is commensurate with what is urgently needed in this new field of QIST. Additionally, some educators noted that the skills needed for QIST are broad enough that individuals can pivot to other fields if they preferred not to continue working in quantum technologies.

For example, Professor Milan emphasized that those considering entering QIST need not worry about being stuck once they enter the field and can take comfort in being able to pivot to another field if they decided to at a later time since ``both the understanding of what quantum computing is about and also the skill set you need to develop to work with that, those two things are transferable to any other field.''

Professor Lyric emphasized that many non-physicists are not realizing that their backgrounds and expertise are directly valuable in QIST saying, ``I would say that if someone's background is in computer science, they probably have more tools than they realize to jump into quantum computing, especially if you start to think about circuit-based approaches. If you abstract away that physical instantiation of quantum information and how it actually happens, then I think that the algebra almost certainly was a course that people took at some stage. So, realizing and recalling that body of knowledge and applying it to this [QIST] field [would help]." They continued, ``It may actually seem less daunting if they took a [QIST] course...There are a lot of different entry points, and the way in which it is presented is very important. So, if you're a computer scientist and you want to jump into this field, your pathways should probably be different than if you're in chemistry or materials science or electrical engineering...But of course, it means that more pathways need to be developed. But I feel that it's actually very important to do this. I personally interact with a lot of people whose expertise is in materials science. I'm not a materials scientist by training [but] how I learn [about materials science] is, a lot of it I picked up basically and extended from my solid-state physics background. But that's not enough to really understand deeply what people in materials science know but that's also why I like to do interdisciplinary research [in quantum with non-physicists].''

Professor Lyric felt that it is important to bring more non-physicists into the field of QIST, e.g., by explicitly pointing out to them that their current knowledge and skills may already align with the needs of the interdisciplinary QIST field. They provided an example, explaining that they co-edited a theme issue to introduce material scientists to QIST related research opportunities and how their existing expertise can be useful for this discipline, ``that was the point of this theme issue. It was to really point out all the opportunities in materials science for quantum future technologies and really expose the community to these opportunities, that this research has a lot of really interesting problems [and opportunities]. For some of these materials, people may be the world's expert on classical applications. I mean, if somebody points out, hey, you know what? This [material you have already been working on for other applications] is a really interesting quantum material, and here's why. Then that person, if they feel bold, can jump into the [QIST] field knowing what they know and saying, `I'm going to dive into this, and I'm going to team up with somebody who's an expert in quantum computing and together we're going to try to make advances [in materials] or do something that could be impactful'. So, I think that that's one way in which people can jump in. This is the pathway that comes from materials science. But it [pathway to enter QIST] would be different for chemistry. It would be different for electrical engineering ..."

Professor Lyric continued, ``[So] you would be pointing out all the ways in which what they're [already] good at is important for the [QIST] field, this new field...that's being developed as we speak. I think that to me, that's a more fruitful way, a welcoming way of introducing this field rather than say you either have [taken] quantum mechanics [courses like those in the physics department] or you don't have quantum mechanics [and if you don't, you do not belong]. It's not like that. First, we can't afford to do that.  But the reason why we're sort of in an era [in QIST] where it is dominated by physicists is because we're in the physical era [of quantum technologies], where the physics of the material that make up these quantum bits and quantum gates actually matters, and these things are limiting us. It's also materials science and chemistry and electrical engineering...Maybe our questions [in QIST] are most familiar to physicists...I've been thinking about closely related things for many decades...[To understand what I am saying, I want you to think of an] example of somebody who's been working on a material for 25 years, and then somebody realizes, hey, that material is a great quantum material. You didn't develop it for that purpose, but it is [useful as quantum material] and [someone says] come over here and let's see if we can develop this for that purpose. So, I think that...this big [QIST] field needs to be welcoming of the expertise of all these related STEM communities to solve these really hard problems [in QIST].''

Professor Lyric stressed that it is also important to bust the myth that getting a Ph.D. in quantum physics is the only way to get involved in QIST. They said, ``I think that there are many, many paths and many ways to get involved. These quantum technologies are very complex, and they necessarily involve that you gotta be in STEM, more or less, unless you're going to be in marketing or communications about the [quantum] field. I would say that there are many opportunities to make meaningful contributions to the large-scale goals [of QIST] whether you're in materials science or in computer science and the traditional fields - physics, chemistry, electrical engineering. There are big, hard problems [related to quantum], big enough to really build a career out of, all of those facets of this really hard challenge. So, I think that [the notion that the quantum computing field is only for some scientists], that's a misunderstanding that maybe comes from the way in which it's packaged to people...Quantum computing-it sounds like it's got quantum, and it's got computing in it. So, [it seems] you gotta either be in quantum or you have to be a computer science major. Maybe that's part of the issue. But again, the quantum aspect of it is something that I think not just physicists, but physics education researchers have an opportunity to clarify." 

Professor Lyric reflected further saying, ``going back to the original question, it's not just about [more people from non-physics backgrounds] being in [QIST] business. In some sense, [who is in this field] depends on the age level [you start learning quantum] or the level of professional development including where you start and where you get your Ph.D. in an area, [when and to what level you] learned quantum mechanics. Let's say, that is kind of a divide. You either did or didn't learn quantum mechanics. Does that mean that you can't [learn relevant things in quantum later]? Or up to what level did you learn it [previously]? That may feel like a barrier that you can't penetrate. And I disagree that that's a real barrier. But maybe a different set of tools would need to be developed that would be more specific to the backgrounds that people have if they want to enter a field [like QIST] at a more advanced stage of their career. They want to join this area [later in their careers]. What do they need to know? And what's the best way for them to learn it? I think these are all important questions that I think are tied in with the need for workforce development of people who are going to actually do research in this growing [QIST] area.''

Professor River noted that either physics students entering quantum industries will require additional training in engineering aspects of QIST, or more non-physicists will have to be trained to participate in this growing field. They elaborated, ``I don't know what's easier: a physics student spending multiple years tackling engineering problems or an engineer maybe spending a year learning quantum mechanics. I do get a sense maybe it's changing over time. We're going to see more non-physicists learn about quantum because really, I don't think it's that difficult. You need time investment but it's not something where I need to devote my life, whole schooling to [if I am a non-physicist]. Again, I don't talk to a lot of them so I can't speak to why engineering or computer sciences are reluctant to get into the quantum efforts.''

\subsubsection{Role of Non-Physicists in Historical Development and Timing Considerations}

Some educators pointed out that the historical developments in QIST have already included major contributions from non-physicists. One educator also expressed uncertainty about whether the field currently needs more non-physicists than those already involved. 

For example, Professor Radiant reflected broadly on the field of QIST currently being dominated by physicists saying, ``maybe that's not a bad thing. Maybe it's natural that as technology improves, it will bring in more people [from non-physics backgrounds] slowly. Maybe if you got too many [non-physicists] now, it would be too many. Maybe at this point, there are not enough problems [for non-physicists] to solve.'' They compared the situation in QIST with that in AI saying, ``I would use the example of AI because in the 1980s...there was a lot of activity in AI theory. The computer scientists and the students were very excited about it. They came up with all these algorithms like neural networks and so on, and there was probably a fairly small percentage, but they were mathematically minded people. They were teaching students courses on AI and they expected it would happen fairly quickly but it didn't. I think the simple reason [AI did not advance quickly] is that the computers were not powerful enough at the time. So, you had to wait...20 or 30 years until the computers became much more powerful. Then they started implementing the algorithms that had actually been lying there for 20 years. So, I would say that the computer scientists who were ahead of the game...were making very important contributions to worldwide knowledge and science research."

They further elaborated on the timeline of advances in QIST, ``So [when and how people decide to participate in a field], it just depends on what the people want to do. If they want to interact with companies now and get consulting jobs and build databases and stuff or build compilers that can be used in science research now, that's their choice; but there should be a small fraction of computer scientists [in quantum computing] like there were in the AI days 30 years ago, who just want to do this [quantum research] because of scientific curiosity and to be in it in the early stages of the field and make a big contribution that other people aren't thinking about. Once it [QIST] grows and gets really big, maybe it's harder to make an impact. But if you get in on the early stages [of QIST], maybe you can make more of an impact even if nobody cares about it [meaning QIST is not impacting people's everyday life] right now.''

Professor Radiant believed that the current dominance of physicists in QIST, compared e.g., to computer scientists and engineers, is because we are still in the early part of the second quantum revolution. They contemplated on the historical development of the interdisciplinary field of QIST broadly since the 1980's, stressing that both physicists and non-physicists, e.g., computer scientists have already contributed greatly to get QIST to this point. Professor Radiant discussed these issues by first reminiscing how they got involved in QIST saying, ``even though I saw Richard Feynman speak on that in one of his famous lectures \cite{simulationfeynman} in the early eighties, I had no idea what he was talking about and so I just went back and kept working on quantum optics, 
[while] it [QIST] has slowly developed into a field with practical applications driven initially by, I would say, quantum key distribution. Our friend Charlie Bennett \cite{bennett1984} was the first person to do that [showed a practical application of QIST in quantum key distribution]. But it still seemed just like curiosity, a strange thing about physics. And then, of course, we had the Bell's inequality experiments \cite{bell1,bell2,bell3} done in 1970s, eighties and later in the nineties but that was mostly under the radar.''  Professor Radiant then emphasized the contributions of Shor, a computer scientist, saying, ``it was only after Shor's algorithm \cite{shorcode,shor1994} came out in [the mid-nineties] that people started to realize, okay, there's something practical here you could do. Then Bennett and others [including those from non-physics backgrounds] kept proving theorems about quantum communication security and using entanglement to do teleportation \cite{teleportation}. Then people realized that teleportation could be used inside of a quantum computer, and then people started talking about quantum sensing...So eventually it grew up organically, very slowly. And my [research] world gradually transitioned into this [QIST] field, as everybody else's did.'' Professor Radiant emphasized that the second quantum revolution is playing a central role in their own research now and they are already working with non-physicists as part of a national engineering research center funded by the US government saying, ``Now a lot of my work is motivated or even justified by this new practical field of quantum engineering [involving both physicists and non-physicists]"
 
Reflecting further on the historical developments in QIST that requires the expertise of both physicists and non-physicists, Professor Radiant said, ``There was no one big breakthrough, so I wouldn't say Shor's algorithm [was the only thing that led to advances in QIST research but Shor's algorithm] was just a stimulus that got the funding agencies in the government, especially the National Security Agency to start putting money into it".  Looking back at the developments in quantum sensing, quantum communication and quantum computing driven by contributions from both physicists and non-physicists they said, ``People thought quantum sensing would be the so-called low-hanging fruit that would be easy to build. Well, there are a lot of sensors that we call quantum 1.0, which are based on previous quantum technologies like atomic clocks, lasers and of course [classical] computers. And there are a lot of enhancements that have been made in those. But so far, nobody has really built a practical quantum sensing system that really uses entanglement in a really major way. The best example I can think of actually is for advances we could call ``quantum inspired", that's using magnetic fields to image the brain. So, there are these groups at hospitals in the UK and US that are building brain images based on magnetic field sensing. And what they do is amazing! [For imaging the brain], they have a helmet they put on the patient, and inside a helmet there are maybe 64 very tiny atomic rubidium vapor cells.'' Professor Radiant thought that both non-physicists and physicists were being hired to make progress in sensing technologies saying, ``[some people were] jumping on to this [QIST] bandwagon to try to develop or at least to hire other people to develop systems, hoping that there will be some really huge quantum advantages coming soon in sensing like radar and remote sensing. But it turns out to be harder than people thought to make quantum sensing work, whereas it's turning out to not be easier, but more realistic to make [the NISQ era] quantum computers work...And quantum communication [involving both non-physicists and physicists] is also moving slowly again because of the difficulty of distributing entanglement over long distances. So, quantum computers have the advantage that all the entanglement is in a very small volume and that's easier to control. But if you want to entangle across...kilometers that's much more difficult to control.''

Discussing contributions of physicists and non-physicists in quantum computing, Professor Radiant said, ``I think we have quantum advantage already in simulation [regardless of] whether it's useful or not [at this point]...I think Feynman was way ahead of his time...His reasoning was that if we could do these manipulations in Hilbert Space, we could do something...[But] then we had to wait...for technology to catch up. And people like David Wineland \cite{wineland}...developed these ion traps, and they were building those to build atomic clocks at one-point.
Then after [computer scientist] Peter Shor came along, other theorists made proposals explaining how you could implement a two-bit quantum gate using trapped ions...Theorists said that you can do this and it is really important. So they did it and then later Wineland got the Nobel Prize for...how to manipulate and control  
the quantum states of single atoms. So, it [QIST progress] is many different incremental steps [involving both physicists and non-physicists]. And then...also the theory was improved [and] people like [computer scientist] Shor came up with error correction codes. There was a lot of communication and talk between computer scientists and physicists, [for example], David Deutsch \cite{deutsch} and Charlie Bennett and all those people \cite{deutsch,woottersclassicalquantumchannel,teleportation,woottersqkd}. So, it is really many different directions and many different small steps [by physicists and non-physicists both] that have finally led to the final [form of QIST that we see at the current time].'' 

Thus, while Professor Radiant reflected broadly on how QIST has historically progressed and emphasized the contributions of both physicists and non-physicists in advancing the field and bringing it to the current state, they did not necessarily think that there is an imminent need to attract more non-physicists to be involved in the field than are naturally attracted to it because they need to be interested in foundational research.

Professor River also emphasized the historical contributions of non-physicists, saying that already in QIST, ``there are some very famous people that are from those [non-physics] fields [computer science, engineering, etc.]".

\subsubsection{Need for Interdisciplinary Collaboration}

Some quantum educators pointed out that there is already an urgent need for collaboration with researchers from different disciplines, and some drew examples from their own research to emphasize this need.

For example, Professor Haven, who works on building a quantum computer using superconducting qubits, believed that research collaboration involving more non-physicists working alongside physicists would be extremely valuable at the current time. They said, ``I think we, for instance, have pushed the Q [factor] \cite{qfactor} of qubits...by a factor of 10,000 since I started graduate school...At that level we, for instance, are probing materials properties of our substrates of our material systems really, really, deeply. That is an area where, in silicon [based technology], that would be the preserve of the people like [those] at Intel [semiconductor company]...We welcome [more non-physics] people who want to do that kind of research with us. But it's not right now a multibillion-dollar industry...So, I don't know what causes them [non-physicists] to enter or not enter into the field. But there is definitely work [of non-physicists to be done] now and probably we need that work done before we can have the really good fault tolerant large-scale machines [quantum computer]." They continued, ``So, it's not that we can build the first large scale machine out of vacuum tubes, and then people can clean it up and make it transistors. It's more like to get to the vacuum tubes, to get to the first working computers, we definitely need these external ideas [from non-physicists].  Error correction itself is not a physics idea. It's really a computer science, information theory idea...This [group of researchers involved in error correction] can be mathematicians or a mixture of mathematicians and computer scientists. That's already a place where I don't think physicists are really in the driver's seat. We [physicists] are building the systems [for example, figuring out ways to build quantum computers using different qubit architectures]. But the ideas [for error correction] come from a traditional computer science background. I think we are starting to see more and more motion even in [my university regarding non-physicists' involvement]. We have now several quantum colleagues in the engineering school, and we have quantum colleagues in [computer science]. [Even at other universities], we see them hiring in computer science and we see that at federally funded center[s]. They have now teams looking at quantum algorithms [in some universities]. So, I think it [the work of non-physicists] is really starting to spread. How fast or how far it spreads, I don't really know, but I do know that we need these sorts of [efforts from more non-physicists in QIST to advance the field].''

Professor Haven also reflected on the important role of non-physicists in interdisciplinary teams like theirs saying, ``You need to work in a [university] lab like mine to learn how to build a [superconducting] qubit. It's just the only kind of place where we know these superconducting systems well enough to build them. It's not that there's an industry close by that you can convert people from. It's a really strange technique. I think these ion traps and neutral atom systems [for quantum computing] again are very sort of specific small crafts that don't link to an industry where you can bring in already trained people and plug them in. But to build functional [quantum computing] devices...we need more computer science. We need more materials science." They elaborated, ``The problem we're having is, the better the systems get, the less easy it is to replace one component [of the quantum computer] without breaking something else. So, one good thing that's, for instance, happening in our quantum center...is that we have teams that have enough different kinds of backgrounds on it that we can start to get a picture of---here are the 10 moving parts, if I take this out-what does it do to the computation? What does it do to the physics? what does it do to the lifetime?...If you try to introduce something new, [e.g.,] a materials scientist says, here's a new superconductor, but it breaks 5 other things about the [quantum computing] system, it doesn't work. It's one reason it's very hard to make progress [in building a quantum computer], even though we're still making [progress]. It [the problem] is that you have to do everything as well as you have been doing and then take something broken and move it up. But you can't sacrifice anywhere else. And that kind of joint complicated optimization [in building a quantum computer] requires that the computer scientists be willing to tweak the algorithm and the material scientists be willing to pick materials that are compatible with what you already have, and the physicists have to go find out what part of nature is getting in your way.''

Professor Milan acknowledged that currently both physicists and non-physicists are contributing to their own QIST research saying, ``I can see that some problems that we're dealing with are more physicsy and some that are more on the engineering side. So, for example, the fabrication of a new device that will do things more efficiently, even though there was some other device that did that before, but with very low efficiency, I think that's more of an engineering challenge. Whereas some new physics in how multiple color centers [one type of system this researcher is interested in for QIST applications] couple to a cavity, that is much more physics than engineering. So, I can see how all of these disciplines can contribute [to QIST at this time] and there are others like material science, at least in just our [current] research, so that was engineering and physics and material science and then also some computer science.''

Professor Sage, who works on semiconducting qubits, provided a thoughtful personalized response saying, ``to encourage interdisciplinarity, it really helps to have a problem that needs a solution. Right now, we are facing one of those — we have an idea for improving the qubits, but we don't know if the structure can be grown. So, you really need to have a good relationship with materials scientists".  Professor Sage drew upon their own research in the field of quantum computing and reflected on the physics and engineering challenges that require interdisciplinary efforts for making progress in quantum technologies, ``I think now that the field has bifurcated into the big efforts developing large scale quantum processors and then the cottage industry … we can use the qubits to really understand the materials. So, [what] I've gone [into] much more in the last few years, it's using qubits to characterize materials as opposed to trying to make qubits because we know how to make qubits more or less. And this is relevant to making better qubits because, again, if you can understand what the noise sources are, that's an important step in being able to improve the gate fidelities and reproducibilities, etc." They described that for their qubit, they are researching how to design structure to make the yields 99\% saying, ``With the students [you work with], typically, what you do is, you try to put it in that context of oh, we're trying to make a quantum computer … here's the thing that goes wrong. And we want to understand how often it goes wrong, how do we fix it, etc. Then you tend to motivate it [to students working with you] by this overall goal of the 99\% yield that you're trying to achieve. But then there's a very specific project that's actually quite physicsy in terms of what they actually do". They emphasized the importance of being able to address both the physics and engineering aspects of the challenges in developing useful quantum technologies by an interdisciplinary team, ``if you're going to scale up, you have to understand how to make it so that every qubit works alright and again, one of the things you can do is use the qubit itself to figure out, map out, where the problems are and then understanding the physics of when you're going to get problems and how to make it so you minimize the problems and you get the yield as high up as possible and what the limits are, etc." 

Professor Sage stated that there may be a point when physics aspects are well understood and engineering aspects of quantum technologies become dominant, ``at some level, it's not completely clear when it turns into engineering to the point where something that you're not expecting is going on and that's the source of your problems". However, Professor Sage believed that there is still a lot of physics to be done alongside engineering efforts, making interdisciplinary teams essential. They stated that at least for them ``discovering unexpected physics is what makes it interesting and at a certain point I think it will get to the point where...how to make the individual pieces is really well understood and then it's more of an engineering problem of not screwing it up".

Professor Sage also reflected on how it took their entire interdisciplinary research team a long time to work synergistically from the idea of a qubit to actually having a semiconducting qubit, while tackling the physics and engineering challenges along the way. They described the continuous trial and error in the process saying, ``It started with nothing. Then, to build it step by step, and to see that you can do this, again you had to figure out what [to do]. So, you would try it. It didn't work. You had to figure out what went wrong. A lot of the work was like, is it possible to work in this material system and make qubits out of it?" They explained that when something did not work at first, it was unclear to them whether there was some problem to be fixed, or that the idea itself was flawed. However, realizing that the idea can actually be successful makes all the work very exciting. 
They continued, ``To get from an idea to a qubit was more than 10 years, and that's something that was amazing that there was…the funding where you laid out a roadmap of what you were going to do, and of course it didn't go as quickly as you were hoping it would go, but they [funders] could see that you were on track and doing [work] and this consistent funding enabled a lot of different technologies to develop to the point where there are now a lot of different technologies that people can [use to] make qubits." Professor Sage also emphasized the physics and engineering challenges that require interdisciplinary solutions, such as the iterative process of attempting to build a qubit and how both the underlying physics and experimental approaches were necessary,
``let's try to make a qubit. Oh, that didn't work. What went wrong?...And what experiments do you do to figure out what's going wrong? And then, once you figure out what's going wrong, can you fix it?"

Professor Sage's reflections made the importance of an interdisciplinary team with expertise in diverse areas in making quantum devices clear, ``At a certain point, we had qubits, and we were all excited, of course. But to make a useful quantum computer, the qubits have to be very high quality. They have to have what's called very high fidelity. So, they can't leak out to the environment. You have to be able to control them very accurately and so they have to be not decohering, but on the other hand, controllable. It turns out that both things are really hard. So basically, what happens then is, you do something, and you realize that they're not accurate enough. Again, what's the problem? Diagnose the problem. What physics is underlying the problem? And again, because you have the design of the qubits, it's not just like, here's physics, what are the numbers describing it, but like, can you make devices that somehow mitigate the problems?"  Professor Sage also gave examples of contributions related to the design of qubits that can help quantum computing without improving the materials further or reducing the noise saying, ``there are so many different dimensions because there's the understanding [of] what the problems are but then sometimes you can change the design of what you're doing, you can change the material stack. There are just so many different directions you can go to try to make it [qubit] work better." These reflections underscore the need for interdisciplinary teams involving both physicists and non-physicists to accelerate QIST.

\subsection{Q3: What are your thoughts on ways to foster and manage university-industry partnerships for innovation in QIST?}

\begin{figure}
    \centering
    \includegraphics[width=\linewidth]{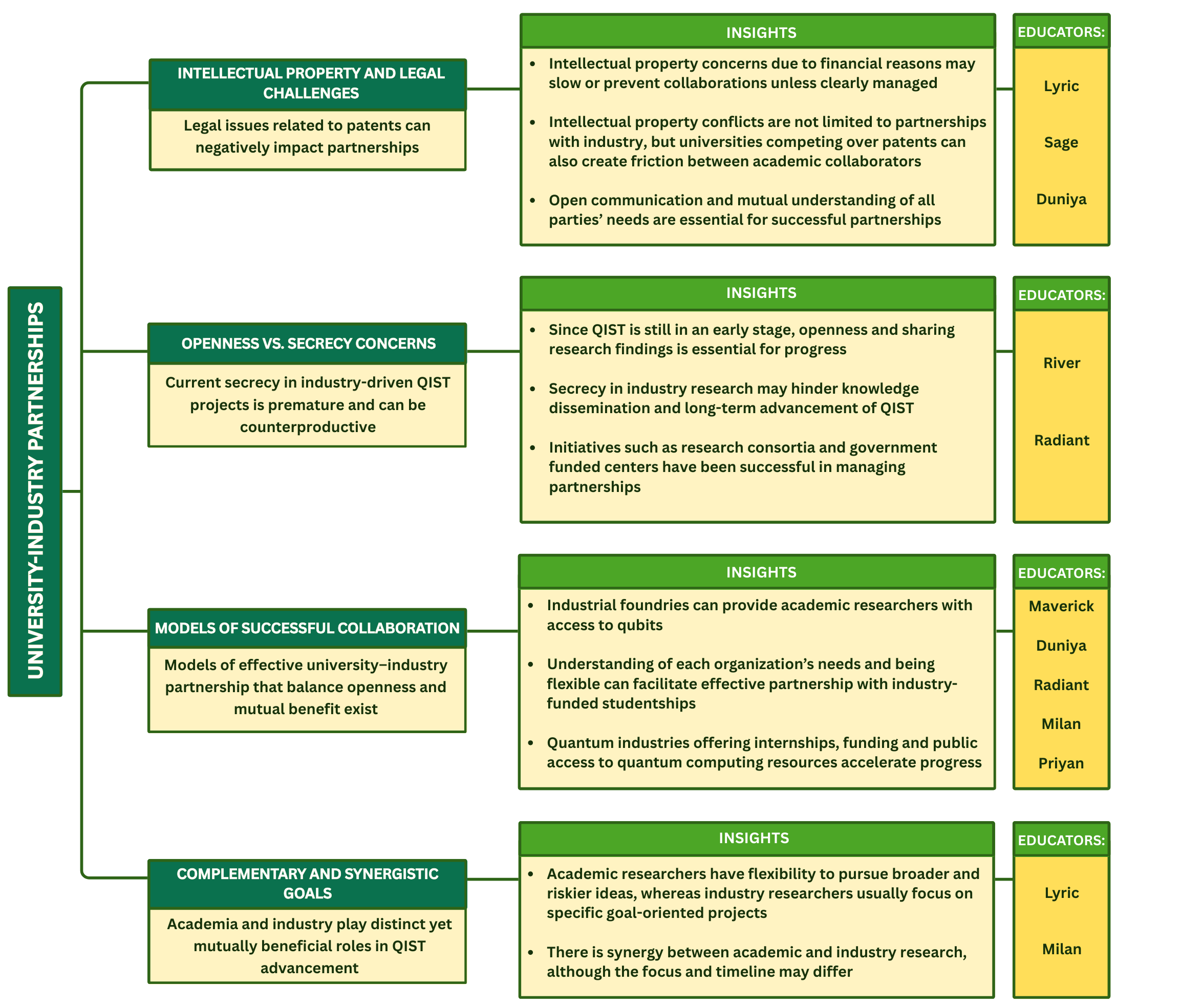}
    \caption{Common codes and insights of educators in response to university-industry partnerships.}
    \label{fig:RQ3}
\end{figure}

Educators recognized that there are intellectual property issues and secrecy concerns that can be impediments to productive university-industry partnerships. However, several educators stressed that figuring out ways around these hurdles is possible, and some outlined what successful collaborations may entail. Additionally, some educators felt that the complementary goals of academic researchers, who are free to explore risky or long term goals, vs. industry researchers, who are more likely to follow a set agenda, can be productive for the overall long-term advancement of QIST. Quantum educators' thoughts related to university-industry partnership were categorized under the following codes: ``Intellectual property and legal challenges", ``Openness vs. secrecy concerns", ``Models of successful collaboration" and ``Complementary and synergistic goals". The insights provided under each of these categories, as well as educators who mentioned them are summarized in Figure \ref{fig:RQ3}.

\subsubsection{Intellectual Property and Legal Challenges}

The fact that intellectual property (IP) issues can negatively impact partnerships between two institutions was mentioned by some of the quantum educators. Careful consideration of these issues and dealing with them is necessary so that progress in QIST is not thwarted. The educators pointed out that since IP issues have underlying financial concerns, without a clear plan for partnership, there could be legal challenges impeding collaboration, e.g., between universities and industries.

For example, Professor Lyric reflected on the challenges in fostering university and industry partnership as two large organizations focusing on their own financial interests. They explained,
``if you think about a researcher wanting to do something with a company, that researcher is part of a university. It is huge and there are rules for how you can and cannot interact [with industry]. The same is true on the company side. And again, with sharing patents, [an important issue is] who has intellectual property if some discovery is made if there's something that's joint. Everybody, like the lawyers...want to have it in writing [that] it's their thing first. These are very complicated, [since these issues] start involving money." They continued, ``the reason why people do this is because every once in a while, there's a jackpot; and the jackpot is hundreds of millions of dollars or billions of dollars in revenue for a university or a company. So, nobody wants to be the fool that gave away the rights to some discovery for whatever reason. So, this is the backdrop for all of this...[and why] there are lawyers...there are a lot of potential financial implications for some of these discoveries that may be happening in a university setting, in a company setting and so forth. And so, there's a reason why, even if you think you're just tinkering with something, and you want to work with a company, suddenly, the lawyers come in and then freeze everything''. They added, ``I'm not saying that it [meaningful university and industry partnership] doesn't happen. It's just that it's complicated and it's complicated by this factor. So, to do it in an ad hoc way, it is not trivial to just say, oh, let's work between academia and industry. Usually there are consortia that help manage [these issues]. They'll say, this is all open and anybody can use your patent within this pool, for example.''

Professor Sage started their reflection by emphasizing that ``the intellectual property aspects are really very troubling" but ``it's not just with industry".  They gave an example of how these types of [IP and patent] issues between two universities became challenging for them to navigate during a sabbatical year at another university. They added, ``that's been actually one thing that's made things much more unpleasant over the last several years, once it [patent issue in QIST] became a thing...it's not just when you talk to a company, but now the universities are fighting over these patent rights, and again, in a way that makes absolutely no sense." These issues could also adversely affect collaborations involving students across universities and industries.

Reflecting on fostering productive university-industry partnership, Professor Duniya noted that their collaborative efforts with industries have gone very well because they always carefully consider the needs of both organizations. They said, ``I think the challenge is a little bit different in different universities. There are some universities that don't get tied up in these IP [intellectual property] discussions at all, and there are some that get completely wound up surrounding IP discussions and how to do these things. I think it requires a lot more communication between people in industry, academics, and university administration, and a greater appreciation for why it is that industry needs patents, and why it is that academics might not need patents to the same extent. So, we have successful agreements [and] we have a patent we submitted with a company". 

\subsubsection{Openness vs. Secrecy Concerns}

Some educators emphasized that quantum technologies are not yet so advanced that companies should restrict publication of research in 
journals where anyone can access it,
and that such issues are unfortunate impediments to university-industry partnership.

For example, Professor River believed that currently we are not at an advanced stage of technology development in QIST that justifies industries being secretive about their research findings. They explained, ``what is certainly something that is a little annoying is that as certain companies invest into quantum technology, they're also making certain advances no longer public." They provided an example of their mutual project with collaborators in a company, and how that brings ``minor hurdles" such as signing disclosure forms, waiting for approvals for commercialization and getting patents, and confidentiality.
They believed that as a theorist, these issues may not be as pronounced saying, ``the problems are worse for experimentalists because they feel like they might actually have something. It's not just [redacted] company [where I have collaborators that has these issues] but all the big companies. So, this [type of secrecy] is something that companies are jumping on, there are certain advances that are no longer being made public." They continued, ``And I'm not sure we are at the stage of research yet where that type of competition is healthy, because I feel like we might still be in the infancy stage, where it benefits everyone to still keep sharing openly a lot of data before we get to the stage where, okay, this is now industry secret...there's that fear that all these big companies might decide this isn't the way to go. Hypothetically, if [redacted company] decides to shut down its QIST division, then the knowledge accumulated [about QIST research] may never be disseminated to the public and be lost.''

Professor Radiant reflected on the contributions of companies like Bell Labs, which hired many physicists and engineers with Ph.Ds., in spurring science and technology innovations several decades ago and whether its absence is detrimental to progress in QIST. They said, ``I think Bell Labs was great. It had its time and it was a real shame when it broke up. But times change and now...there are a lot of small companies and a lot of the company work is being held more privately. It is true that Bell Labs did pursue patents...They were very strict about that, but at the same time, they did publish much of their work in open literature. They had postdocs going in and out, and so on. There was more connection." They continued, ``Now I think...industry is very eager, of course, to hire PhD's from our groups but then, once they get hired, they kind of disappear and we don't really get to know what they're working on. That is true. There's nothing that really replaces Bell Labs. That's why I guess the National Science Foundation (NSF)...they funded all these large centers on NQI (National Quantum Initiative) funding to try to bring people together and they tried to also encourage these NSF centers to involve companies. So there has been some success there." They continued to provide an example from their own experience, ``I'm part of [redacted] quantum center. I'm funded partly by them and that center has probably 30 or 40 companies that are corporate members. They have different levels of interaction, but we've had a lot of scientific exchange. In fact, I even just co-authored a [specific journal] paper with one scientist from [redacted] company. We're getting the students to get to know these industry scientists...There's not one monster that's Bell Labs. It's a whole bunch of different companies that are trying to get involved with university people. So, I don't see it as a major problem, I think it's just different now.''

Professor Radiant added that they were unsure whether no one company like Bell Labs having a monopoly is hurting progress in QIST saying, ``Well, of course, the history is that Bell Labs had a monopoly on long-distance phone services and that's why they were wealthy. That's why they could afford to build this kind of luxurious university-style [company]. It was a monopoly, and politicians decided they didn't want to have that monopoly funded anymore by the public good. The analogy would be maybe...if Meta were the only company that's allowed to use the Internet, and then they get all the money. Then they can do whatever they want and maybe they would be more open [to sharing information about progress in research instead of keeping it a secret]...if they didn't have competition. But now, of course, there are many people in the government who think competition is the way to go. So, they killed Bell labs. And now we have competition [in industry since there is no monopoly]. I don't know if it's good or bad [for research], but that's just the way the US system works.''

\subsubsection{Models of Successful Collaboration}

Some quantum educators discussed already successful collaborative models for university-industry partnerships that could be valuable for others to follow if they are interested in productive collaborations.

For example, while Professor Maverick acknowledged the challenges in such partnerships, they were overall optimistic about the possibility of working out productive approaches for collaboration between researchers at the universities and industries. They reflected, ``...it is possible [to foster meaningful university and industry partnerships]. People have to get creative around that. So, [earlier] I was mentioning that qubit collaboratory where Intel is giving foundry made chips to academic groups; that's one possibility [for such a partnership]. Many, many large corporations have set up collaborations with academics. Microsoft is one example...Then there are startup companies that have a whole range of ways in which they interact with the university that they spin off from. There are startups that are born out of the university. But then they keep a very good separation between what's the company and what's the university side. There are other startups that are basically one blur of university and company..." They continued, ``So, it [university and industry partnership] is one thing [in which] there's no right or wrong way to do it. We'll just have to see which one works best, and different models may work best in different circumstances. It's all very delicate, mind you. None of these is easy and there are plenty of land mines, be it for intellectual property agreements, be it for the general degree of confidentiality besides patents and stuff,
just the general degree of confidentiality that is required once you're working on a commercial activity. It chokes free flow of information in and out of the company. And it can be a problem, but also not. There are classified programs where that degree of secrecy is even in the extreme, even much more than commercial operations. And yet they are able to make progress. So, there's a whole range of things. There's no right or wrong answer.''

Professor Duniya described the successful partnership at their university saying, ``a lot of it comes down to, is there a company that wants to essentially immediately license the patents, to the extent they are willing to pay for the cost of filing the patent and if so, then we will definitely go ahead. I think that's a healthy way of doing it. We often don't need patents. But if we have industry partners and they need patents to protect their route to commercialization of the technology and for various other business reasons, I think that understanding then becomes important." They continued, ``We have also three industry funded studentships. All of them involve a certain amount of IP that would be retained by the company, but all of them are for exploration of relatively fundamental things that we don't expect to generate a lot of IP in the course of studentship. They could lead to projects that might generate more IP and therefore more IP related issues later on. But I think we've done well at characterizing where this isn't a major problem and then signing agreements that are not overly protective from the university side when we don't need to be. The key thing we always have in our agreements is making sure that students can publish papers and submit a thesis.''

Professor Radiant reflected, ``Some universities like mine have developed master’s-level industry internship programs as a means to rapidly move students who have completed an undergraduate degree in physics or engineering into the quantum industry. In one model of such a program [that we use], students spend around 15 months, focused on gaining experimental skills needed in the quantum industry, which overlap with skills needed in the nanotechnology industry. Such a program includes initial coursework that emphasizes hands-on lab experience, followed by a paid internship spanning six to nine months at a quantum-related company or in a university laboratory."  

Professor Milan stressed the openness of quantum industries in educating students and how it benefits the entire QIST ecosystem. They noted that quantum industries are interested in collaborating, providing internships to students and funding some academic research saying, ``I definitely saw encouraging situations, whether internships or gap years that students have been able to do in a quantum industry, or how open different quantum companies are to collaborate with academia and not just keep the door closed, or [several companies that want to] fund some of the [academic] research.'' Professor Milan also emphasized the value of companies making their quantum computers available to everyone including those in academia saying, ``It started with IBM \cite{qiskit}, but then everyone followed suit, opening quantum computers to anyone or to some people to apply for academic grants. All that has been very useful. I think, at least, in a way it is unprecedented that such an expensive resource, such a rare resource [quantum computer], is democratized that way. I think it really speeds up the development. So, I'm under the impression that it [university industry partnership in QIST] is going better than in some other fields.'' 

Reflecting on a particular way of viewing successful university-industry partnership, Professor Priyan said, ``the current way [is] that academia feeds industries by training people and sends them to work at [companies like] Google and IBM. That's not a terrible model, right?" They added, ``you need industrial level engineering to get further in these quantum computing things. Academic nano fab, academic facilities are not going to produce a fault tolerant quantum computer, we know that. To me, it seems like right now, they [universities] are training people that go to work [in industries]. If industry had more money to support a larger research infrastructure, that would certainly be better. But we would still be feeding them Ph.D., scientists. Yeah, we need more of them [students prepared for the quantum workforce]".

\subsubsection{Complementary and Synergistic Goals}

Some educators pointed out that university researchers have more flexibility to pursue broader research goals and potentially more freedom to explore more risky projects than industry researchers. This dichotomy can be beneficial for the overall progress of QIST.

For example, Professor Lyric emphasized different yet complementary goals of researchers in academia and industries at this time in the development of QIST. They stressed that those in academia have the flexibility to step back and explore a broader ``phase space" in their research that may not appear to produce immediate advancement in QIST, even though it broadly explores issues that could greatly advance QIST research for the long term. They said, ``We are at a stage where we have lots of really cool ideas and endpoints. We know that there are stars in the sky [big goals to achieve such as building a scalable quantum computer]. That's exciting, but we don't know how to get there yet. That's also not bad; it's really why I would say that the science aspect of this [QIST] has been kind of front and center because we're still trying to develop and understand things. We're not just trying to make something better and so it's an exciting time, I think, for a scientist to be able to try to develop physical foundations." They continued, ``I'm a physicist. So, I'm interested in developing the physical foundation of future technologies, the ones that are stars in the sky-quantum computing, quantum sensing and so forth. I know they [these big goals] are there. I know when I'm moving in that direction, and I know when I'm moving at right angles, and I don't feel like I always have to go straight towards that goal. I feel like I need [to explore the] phase space, I need to be able to move around and find maybe a better way to get there [to the big picture goals]." They concluded, ``That's the flexibility that I think we have in academia to do science, not just take one path and run with it. I think that that's one of the things that distinguishes companies from universities. Companies need to focus on what they want to pursue, and they're going to [tell industry researchers to] go ahead and pursue that with a lot of force and might and resources. In academia we can say, I'm going to put this [aspect of my research] on the side for now [and] I'm going to look at this thing [another research project] because this looks like it might address the problem that we couldn't fix over there. And [unlike industry], we [in academia] have this flexibility to move around and to develop a number of ideas and to draw inspiration from different areas [to make long-term progress]''.

Professor Milan stressed the valuable synergy between the type of research conducted at industries and universities, pointing out that while academia is generally more focused on fundamental research in QIST, industries are also contributing to this research. However, their focus and timeline may differ. They said, ``I think university is where a lot of things [fundamental research related to QIST] can happen. I'm happy to see that so many people are also doing this in industry. That brings much more money in terms of how fast we can move this [research in QIST]. And it's nice to then separate [the types of problems that people at the university and industry should each focus on, even if they are partnering with each other]. [For example], this [particular] problem is not for the university because industry is going to solve this in the next 2 years, so [at the university], let's focus on the next big thing that is too risky for the industry.''

\section{Discussion}

Our findings reveal a nuanced understanding among quantum educators of the complex sociological and organizational challenges facing QIST. These perspectives provide valuable insights for navigating the field's development beyond purely technical considerations. Our theoretical frameworks allow us to understand educators' perspectives not merely as individual opinions but as reflections of their positioning within this sociotechnical system. As boundary spanners between fundamental research and practical applications, between physics and other disciplines, and between academic and industrial contexts, quantum educators occupy a unique vantage point for observing and potentially influencing the field's developmental trajectories. Consistent with our theoretical frameworks, our findings also suggest that sustainable development of QIST requires not just technical advances but successful negotiation of social and organizational challenges. Our findings emphasize the importance of understanding how different stakeholder groups construct meaning around quantum technologies and how educational practices can shape more realistic visions of the field's future that account for the perspectives of diverse stakeholders consistent with our frameworks.

\begin{figure}
    \centering
    \includegraphics[width=0.8\linewidth]{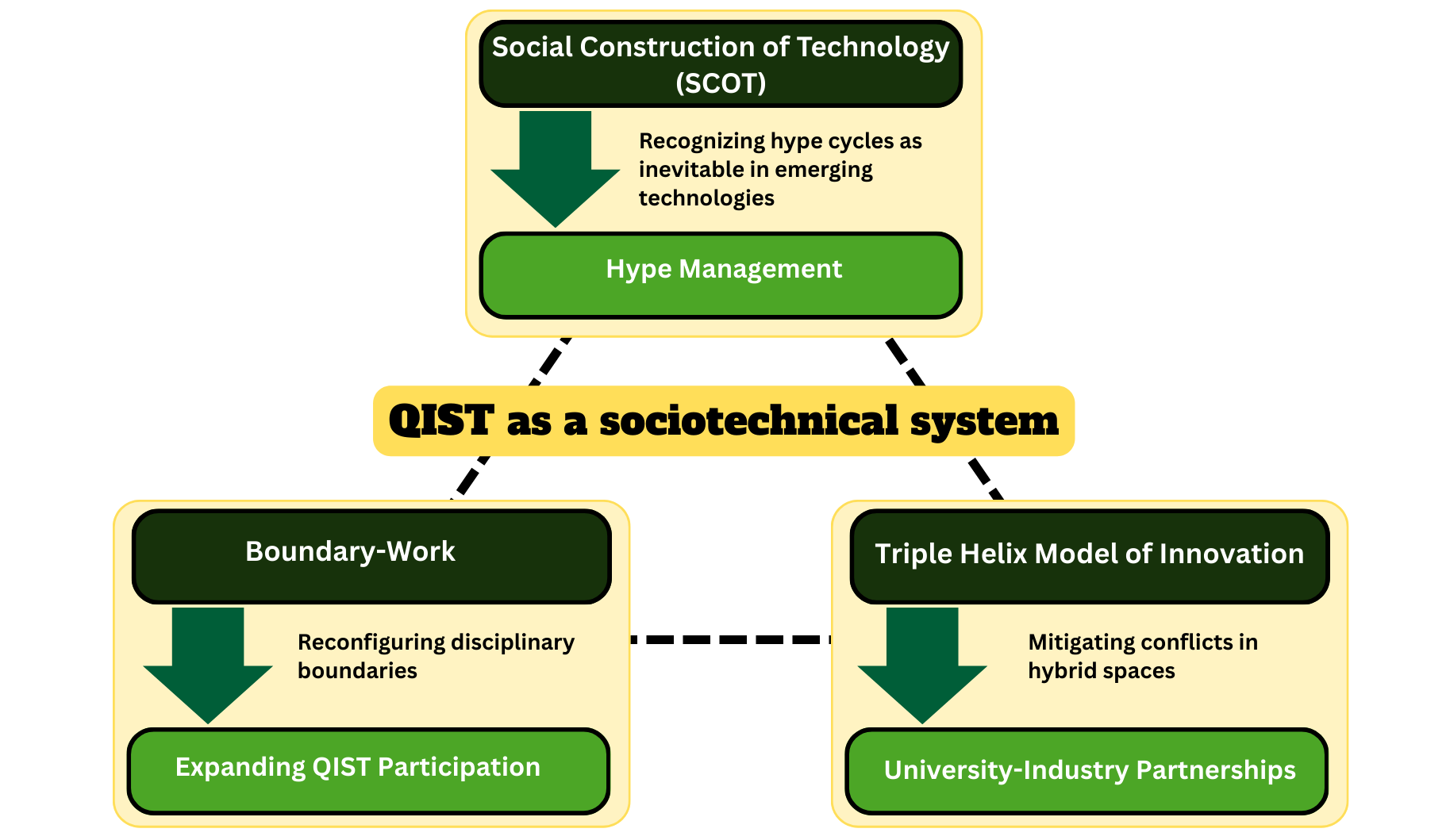}
    \caption{Synergistic frameworks conceptualizing QIST as a sociotechnical system aligned with our research questions.}
    \label{fig:frameworks}
\end{figure}

In the context of QIST, the Social Construction of Technology (SCOT) \cite{bijker1987social,pinch1987social} and hype cycle frameworks \cite{hypecycle,fenn2008hype,brown2003hope} help contextualize educators' observations about the boom-bust patterns in quantum technologies and strategies they recommended for managing, e.g., stakeholder (such as student, investor and public) expectations. These perspectives suggest that hype is not simply a distortion of technological capabilities but an integral element of how emerging technologies develop. For example, the expectations and narratives surrounding quantum technologies actively shape research priorities, funding decisions, and career choices, thereby influencing the technology's ultimate trajectory. We observe both boundary maintenance \cite{gieryn1983boundary,gieryn1999cultural}, e.g., physics continues to dominate the field and boundary crossing, e.g., the field increasingly requires expertise from computer science, engineering, materials science and other disciplines. Some educators' emphasis on creating multiple entry points and recognizing existing expertise of non-physicists can be viewed as efforts to reconfigure these disciplinary boundaries to enable more inclusive participation to accelerate the progress of QIST further. Furthermore, in QIST, we observe the 
emergence of what has previously been termed ``triple helix spaces" \cite{etzkowitz2000triple,leydesdorff2000triple}, i.e., the need for collaborative environments where academic researchers, industry practitioners and government representatives (e.g., who provide funding and regulatory framework) can work together on shared challenges. The intellectual property concerns and cultural differences between universities and industries highlighted by our interviewed educators reflect the tensions inherent in creating these hybrid spaces while maintaining the distinct missions of each institution. Figure \ref{fig:frameworks} presents a schematic of these synergistic frameworks. Below, we summarize some of the key findings.

\subsection{Managing the Hype: A Double-Edged Sword}

The educators' responses reveal that hype is both a necessity and a potential danger for QIST. While acknowledging that hype attracts funding, talent, and public interest, which are essential ingredients for field growth, they also recognize its potential to create unrealistic expectations that could ultimately harm the field's credibility. Several educators drew parallels with other technologies, particularly artificial intelligence and fusion energy, suggesting that boom-bust cycles may be inevitable in emerging technologies. The prediction of a potential ``quantum winter'' by some educators reflects that sustainable progress requires managing expectations carefully. This perspective is particularly important for educators who must prepare students for careers that may span multiple hype cycles. The distinction between scientific progress and business imperatives emerged as a key theme. As Professor Maverick's anecdote about the bank CEO illustrates, there can be fundamental misalignments between how scientists and business leaders view quantum technologies. This gap presents both a challenge for accurate science communication and an opportunity for educators to serve as bridges between these communities.

\subsection{Breaking Down Disciplinary Barriers}

The responses to questions about interdisciplinary engagement reveal both opportunities and obstacles for broadening participation of non-physicists in QIST. A recurring theme is that many of the barriers are perceptual rather than fundamental, e.g., computer scientists, materials scientists, and engineers often possess more relevant skills for QIST than they realize. Professor Lyric's emphasis on creating multiple entry points and tailored educational pathways reflects a valuable approach for how to draw researchers from other disciplines and make QIST more accessible. In particular, the strategy involving explicitly highlighting how existing expertise of researchers in other fields maps onto quantum challenges of this century represents a welcoming approach that could significantly broaden the talent pool in QIST. The historical perspective provided by Professor Radiant, tracing QIST's development from Feynman's early vision through contributions from both physicists and non-physicists, e.g., computer scientists, underscores that the field has always benefited from interdisciplinary collaboration. However, based upon the reflections of several interviewees, the current dominance of physicists in QIST needs to be supplemented with more researchers from other disciplines at a time when the field needs diverse perspectives to collectively brainstorm and overcome technical challenges.

\subsection{Navigating University-Industry Partnerships}

The discussions on university-industry partnerships reveal a complex landscape of opportunities and challenges. While there was general agreement that such partnerships are essential for accelerating QIST research, translating research into practical applications and giving students exposure to what it is like to work with industrial partners (who might be their future employers), educators identified several persistent obstacles. Intellectual property concerns emerged as a major friction point with different universities charting agreements with industries while recognizing that they have varying levels of flexibility in negotiations. The success stories shared by educators, e.g., professor Duniya, suggest that productive partnerships are possible when both parties understand and respect each other's needs and constraints. Some educators, e.g., Professor Priyan, also underscored that universities will keep preparing and sending their graduating students to industries. The contrast between the current model of university-industry engagement and the historical example of Bell Labs sparked interesting reflections. While some educators reflected on the benefits of the more open research culture of Bell Labs due to its monopoly, they recognized that the current competitive landscape, despite its challenges, may succeed in driving innovation in different ways.

\subsection{Implications for the Quantum Ecosystem}

These findings have several important implications for building a sustainable quantum ecosystem. Educators play a crucial role not just in teaching technical content but in helping students, policymakers, and the public develop realistic expectations about quantum technologies' capabilities and timelines. The need for multiple entry points and tailored pathways suggests that quantum education programs should be designed with diverse student backgrounds in mind, explicitly emphasizing how different disciplinary expertise can contribute to effectively tackling various quantum challenges. Furthermore, universities and companies need to develop partnership models that balance openness with commercial needs. Such models can be particularly helpful for training students and giving them exposure to industry research while they are still finishing their degrees. The success stories shared by educators suggest that creative solutions are possible when both parties prioritize long-term development of quantum technologies. Moreover, preparing students for potential hype cycles and market fluctuations is as important as technical training. However, as Professor Milan noted, the transferable skills developed in quantum research can provide career flexibility.

\section{Summary and Conclusions}

This study captured perspectives from leading quantum educators on three critical challenges facing the quantum information science and technology field: role of hype,
role of non-physicists in QIST and fostering interdisciplinary collaboration, and building effective university-industry partnerships. Our findings reveal that these challenges are deeply interconnected and require thoughtful navigation to ensure the field's sustainable development. One key finding is that while hype attracts essential resources and talent to QIST, it also creates unrealistic expectations that could lead to disillusionment. Educators see their role as helping to calibrate expectations while maintaining enthusiasm for the field's genuine potential. Understanding how QIST has evolved through contributions from multiple disciplines and how other technologies have navigated hype cycles can provide valuable lessons for the field's future development. Another major finding is that many barriers preventing non-physicists from engaging with QIST are perceptual rather than fundamental. Creating multiple entry points and explicitly showing how existing expertise maps onto quantum challenges can significantly help in broadening participation of non-physicists. Another major finding is that university-industry partnerships face persistent challenges around intellectual property, confidentiality, and differing organizational cultures. However, successful models exist when both parties understand and respect each other's needs and are willing to yield along some dimensions that may be more important to the other party. Another central finding that emerges from these interviews is that quantum educators must prepare students not just technically but also for the sociological and economic realities of working in an emerging technology field, including potential boom-bust cycles and the need for interdisciplinary collaboration. An effective university and industry partnership, e.g., of the type described by Professor Duniya, can be valuable for students to get an opportunity to work with industrial partners through industry funded studentships.

These findings suggest that building a sustainable quantum ecosystem requires attention to sociological and organizational factors that are often overlooked in favor of technical challenges. Educators play a crucial role in this process, serving as bridges between disciplines, between academia and industry, and between current capabilities and future possibilities. As QIST continues to evolve, periodic reassessment of these challenges will be essential. The perspectives captured in this study provide a baseline for understanding how the field's leaders view these critical issues at this point in QIST's development. Future research should track how these perspectives evolve and whether the strategies suggested by educators prove effective in building a more robust and sustainable quantum ecosystem.

\vspace*{-.35in}
\section*{Acknowledgments}
We are grateful to all educators who participated in this research. This research is supported by 
the US National Science Foundation Award PHY-2309260.

\section*{Data availability statement}
The data cannot be made publicly available upon publication because they contain sensitive personal information. The data that support the findings of this study are available upon reasonable request from the authors.
\vspace*{-.35in}
\section*{Ethical statement}

This research was carried out in accordance with the principles outlined in the University of Pittsburgh Institutional Review Board (IRB) ethical policy, the Declaration of Helsinki, and local statutory requirements. The educators provided consent for use of the interview data for research and publication, and consent for quotes to be used.

\vspace*{.15in}
\Large{References}
\vspace*{-.35in}
\bibliography{refs}

\end{document}